\documentclass[showpacs, twocolumn]{revtex4-1}
\usepackage{graphicx}
\usepackage{amsmath}

\begin{document}

\title{Collective modes of a one-dimensional trapped Bose gas in the presence of the anomalous density}

\author{Abdel\^{a}ali Boudjem\^{a}a}

\affiliation{Department of Physics, Faculty of Exact Sciences and Informatics, Hassiba Benbouali University of Chlef P.O. Box 151, 02000, Ouled Fares, Chlef, Algeria.}

\email {a.boudjemaa@univ-chlef.dz}

%\date{\today}

\begin{abstract}
We study the collective modes of a one-dimensional (1D) harmonically trapped Bose-Einstein condensate (BEC) in the presence of the anomalous density
using the time-dependent-Hartree-Fock-Bogoliubov (TDHFB) theory.
Within the hydrodynamic equations, we derive analytical expressions for the mode frequencies and the density fluctuations of the anomalous density
which constitutes the minority component at very low temperature and feels an effective external potential exerted by the majority component i.e. the condensate. 
On the other hand, we numerically examine the temperature dependence of the breathing mode oscillations of the condensate at finite temperature in the weak-coupling regime.
At zero temperature, we compare our predictions with available experimental data, theoretical treatments and Monte carlo simulations in all
interaction regimes and the remaining hindrances are emphasized. 
We show that the anomalous correlations have a non-negligible role on the collective modes at both zero and finite temperatures.

%The formation of the solitonic structure in such a mixture is numerically investigated using the TDHFB model.
%We show that the pairing component becomes a system of effectively attractive/repulive atoms leading automatically to the generation of bright/dark solitons.
%This soliton forms a bound state with that of the condensed component in the case of high pairing fluctuations.
\end{abstract}


\maketitle

\section{Introduction} \label{intro}

Ultracold atomic Bose  gases in a 1D geometry \cite {Kitt,Goz} have attracted a great deal of interest as they yield a fascinating physics 
and many surprises not encountered in 2D and 3D. 
The regime of a weakly interacting gas requires the criterion $ \gamma= mg/\hbar^2 n_c \ll 1$ which is the ratio between the interaction energy
and the kinetic energy of the ground state, where $g$ is a coupling characterizes the interaction, $m$ the atomic mass and $n_c$ the 1D condensed density. 
In such a regime the mean-field theory can be safely used and a trapped true BEC is possible.
In the limit $ \gamma\rightarrow \infty $ which corresponds to small density or large interaction, the gas transforms  into a strongly interacting system
and acquires Fermi properties. In this case it is called a gas of impenetrable bosons or gas of Tonks-Girardeau (TG) \cite{Tonk, Gir}.

Bose gases in very elongated traps which feature of 1D condensate, have been studied in different regimes of quantum degeneracy \cite{Shly1,Shly2, Dunj}.
The sum-rule approach \cite{Menot},  time-dependent modified nonlinear Schr\"odinger equation (MNLSE) \cite{Choi},
numerical Monte carlo simulations \cite {Astrak, Zvon},  Hartree-Fock-Bogoliubov (HFB)-based approaches \cite {Zvon, Hu, Xiao} and finite-temperature hydrodynamics \cite {Bouch}
have successfully predicted the collective modes of both weakly and strongly interacting 1D Bose gases at zero and finite temperatures. 
Experimental investigations of the breathing oscillations in 1D ultracold gases have been reported by several groups \cite {Tilm, Haller,Bess}.
The main finding emerging from these explorations is that the breathing mode frequencies as a function of the interaction strength, go through
two crossovers: from the value 2 down to $\sqrt{3} $ and then back to 2, as the system goes from noninteracting
to weakly interacting and then from weakly interacting to the TG regime \cite{Haller}.

%However, the collective modes at finite temperatures remain largely unexplored theoretically despite the first experiment \cite{Tilm} was realized more than one decade ago. 
At nonzero temperatures,  the elementary excitations arising from quantum and thermal fluctuations
are always present \cite {Hald, Shly2,  Mora, Gaza, boudj1}  and may considerably affecte the dynamics of the condensate.
It is well known that at finite temperature, the condensate coexists with the cloud of thermal excitations and the anomalous component.
This latter quantity which, survives naturally in Bose systems, can be interpreted as the density of pair-correlated atoms.
This pairing phase is similar to the BCS phase in superconductors at low temperature proposed by Evans and Imry \cite {Evans} a long time ago. 
In addition, it has been argued that the anomalous density accompanies in an analogous manner the condensate: 
they both arise from the gauge symmetry breaking \cite{boudj2, Yuk}. 
If the condensate density is nonzero, the anomalous average is also finite.
Conversely, when the condensate becomes zero, the anomalous density has also to disappear.
Moreover, in homogeneous 2D Bose systems, the condensate and the anomalous density exist together at zero temperature while they both 
do not survive at any nonzero temperature \cite{boudj1} due to the destruction of the long-range order.
In spatially uniform 1D Bose gases, BEC and the anomalous states do not occur at all temperatures.  
However, the presence of a trapping potential in such systems introduces a finite size of the sample and drastically changes the picture of long-wave
fluctuations of the phase and thus, the phase coherence over the finite sample size is restored.

Certainly, the presence of the anomalous density in an ultracold Bose gas plays a crucial role in its dynamical and thermodynamical properties.
First of all, the anomalous phase is usually of the order of or even larger than the noncondensed density \cite {boudj2, Yuk, Griffin, Griffin1, Burnet, boudj3, boudj4, boudj5}.
So, the exclusion of such a quantity is indeed an unjustified approximation and may render the system unstable \cite{boudj2, Yuk}.
Moreover, the absence of the anomalous density prevents the occurrence of the superfluidity in Bose gases \cite{Yuk, boudj1, boudj3, boudj4}, 
which is natural since both quantities arise from atomic correlations.
The inclusion of the anomalous correlations in a 3D harmonically trapped Bose gas, accounts well for shifts in the lower-lying excitation of the JILA experiment \cite{Burnet, Giorg}.
In attractively interacting BECs, it has been found that the anomalous fluctuation 
causes the condensate to collapse and the 1D soliton to split into two solitonic structures \cite{Eric, Bul,boudj6, boudj7}.

In view of these circumstances it is then interesting to study the properties of a quasi-1D weakly interacting BEC in the presence of the anomalous density.
Obviously, this mixed system is categorically different from that of a pure BEC especially at intermediate temperatures where the anomalous density reaches its maximum 
\cite {boudj2,Burnet}.

The aim of the present work is to investigate the collective modes of both the condensate and the anomalous components in a quasi-1D trapped Bose gas 
at finite temperature utilizing our TDHFB  theory \cite {boudj1, boudj2, boudj4, boudj5,boudj0,  boudj6, boudj7, boudj8, boudj9}. 
The TDHFB is a self-consistent approach describing the dynamics of ultracold Bose gases.
The main feature of this theory is that it preserves the validity of conservation laws, and a gapless spectrum of collective excitations.
The TDHFB approach is based on the time-dependent Balian-V\'er\'eroni (BV) variational principle \cite{Balian} which requires 
the minimization of an action which involves two variational objects: one is related to the observables of interest and the other is akin to a density matrix \cite{Balian}.
The theory is valid for arbitrary interacting Bose systems, whether equilibrium or nonequilibrium, uniform or nonuniform, in the
presence of any external potentials, at any temperature and in any dimension.  

%In the second part, we will investigate the formation of matter-wave solitons in a BEC in the presence of the anomalous fraction.
%The numerial similation of our TDHFB model allows us to show the generation of solitons in the pairing component. 
%These solitons which called solion-pair are reminiscent of pair of optical pulses propagate without absorption. Experimentally, soliton-pair  can be
%made if two lasers are applied to a three-level system, the atoms will be driven to a population trapped state, and a medium that is opaque to a probe laser can, 
%by applying both lasers simultaneously, be made transparent \cite{Alz, Bol, Gry, Bout}.

The rest of the paper is organized as follows. 
In Sec.\ref{flism}, we introduce our TDHFB model which describes selfconsistently  the dynamics of a trapped Bose gas at finite temperature and 
establish its validity criterion. 
We analyze the profiles of the condensed and anomalous densities for different temperatures.
In Sec. \ref{CM}, we use the TDHFB within the hydrodynamic equations to investigate the collective oscillations and the density fluctuations 
of the condensate and the anomalous fraction. 
This quite general approach permits us to obtain an analytical expression for the dispersion relation of the anomalous component in a quasi-1D setting. 
We highlight effects of the anomalous fluctuations and temperature on excitation frequencies.
At zero temperature, we compare our findings for the breathing mode frequencies of the mixed system with recent experimental, theoretical and numerical predictions 
in the entire spectrum of interaction regimes from the non-interacting gas, passing by the mean-field regime, to the TG limit.
%Section.\ref{SM} is devoted to analyze the behavior of solitons in the mixture where we show that a bright/dark soliton can be spontaneously generated in the pairing component
%without any external perturbation or use of the Feshbach resonance.
%Under certain conditions this soliton-pair can form a bound state (soliton-molecule) with the condensed soliton.
%On the other hand, we demonstrate that the so-called pairing instability may lead to split the soliton in the condensed component into several peaks forming a soliton-train.
Our concluding remarks are presented is Sec.\ref{concl}.

\section{TDHFB Formalism} \label{flism}
We consider a weakly interacting Bose gas confined in a highly anisotropic trap 
where the longitudinal and transverse trapping frequencies are $\omega_{x}/\omega \ll1$. 
In such a case, the system can be considered as quasi-1D and, hence, the coupling constants effectively
take their 1D form, namely $g =2\hbar\omega \,a$, where  $a$  is the $s$-wave scattering length.
The TDHFB equations which we choose to employ here constitute a model well suited to this task because it
governs both the dynamics of the condensate and the anomalous density  at finite temperature. 
In this quasi-1D geometry, the TDHFB equations  may be represented  as \cite {boudj1, boudj2, boudj4, boudj5, boudj0, boudj6, boudj7, boudj8, boudj9}:
\begin{subequations}\label{E:td}
\begin{align}
&i\hbar \frac{\partial}{\partial t} \Phi = \left [-\frac{\hbar^2}{2m} \frac{\partial^2}{\partial x^2} +V+g_{ca} n_c+2g\tilde{n} \right]\Phi, \label{E:td1} \\ 
&i\hbar \frac{\partial}{\partial t} \tilde{m} = 4\left[-\frac{\hbar^2}{2m} \frac{\partial^2}{\partial x^2}+V+G (2\tilde {n}+1)+2gn\right]\tilde{m}, \label{E:td2}
\end{align}
\end{subequations}
where $V(x)=m\omega^2 x^2/2$  is the harmonic trapping potential with frequency $\omega$,
$\Phi (x)=\langle \hat\psi (x)\rangle$  is the condensate wavefunction, $n_c(x)=|\Phi(x)|^2$ is the condensed density,
the noncondensed density $\tilde{n}(x)$ and the anomalous density $\tilde{m}(x)$ are identified respectively with 
${\langle \hat\psi^{+} (x) \hat\psi (x)\rangle }-\Phi ^{*} (x) \Phi (x)$ 
and ${\langle \hat\psi{ (x)} \hat\psi (x)\rangle}-\Phi {(x)}\Phi {(x)}$, where $\hat\psi^{+}$ and $\hat\psi$ 
are the boson destruction and creation field operators, respectively. The total density in BEC is defined by $n=n_c+\tilde {n}$. 
The parameter $g_{ca}=g(1+\tilde {m}/\Phi ^2)$ stands for the renormalized coupling constant \cite{boudj5,boudj8, Wrig},
and $G=gg_{ca}/4(g_{ca}-g)$. Consequently, the small parameter of the weakly interacting regime becomes:
\begin{equation}  \label{diltparm}
\gamma'=\gamma \left(1+\frac {\tilde {m}} {n_c} \right) \ll 1,
\end{equation}
indicating that a dilute 1D system requires small anomalous fluctuations.

If one neglects both the noncondensed and the anomalous densities ($\tilde{n} = \tilde{m} = 0$), the set (\ref{E:td}) simplifies to the Gross-Pitaevskii (GP) equation.
For $ \tilde{m} = 0$, the TDHFB equations reduce to the HFB-Popov (HFBP) approach 
which is gapless and coincides with the HF approximation at high temperature. 
Very recently, the HFBP approximation has been applied in a harmonically trapped 1D Bose gas to study the collective modes at finite temperature \cite {Xiao}. 
Note that the TDHFB equations have been discussed by many authors using different methods (see e.g. \cite {Bul, Tim, Stoof, Holl, Cherny} ). 
Moreover, in spirit of the mean-field theory, linearized TDHFB equations have been derived in \cite {Giorg} to study the damping 
and the collective oscillations in the collisionless regime.
Our TDHFB formalism is usually given in the form of nonlocal coupled equations for the condensate order parameter and the single particle density matrix \cite {boudj1, boudj2, boudj0}.
A striking advantage of such  equations would be that they can be solved self-consistently using the exact non-local interaction potential (dipolar, gravitational, etc.).
%However, although the computation with non-local potentials is feasible, is numerically consuming.

The validity criterion of the TDHFB theory (\ref{E:td}), as all HFB-like theories, requires that the temperature should be much smaller than a certain characteristic
temperature  $T_{\phi}$ \cite{Shly1,Hu}
\begin{equation}  \label{Crit}
 T\ll T_{\phi} = \frac{\hbar \omega}{\mu_c} T_d,
\end{equation}
where  $T_d \approx \hbar \omega N$ is the degeneracy temperature (the Boltzmann constant $k_B=1$) and $\mu_c$ is the chemical potential of the condensate. 
At $T\ll  T_{\phi}$, both the density and phase fluctuations are suppressed, and hence, there is a true condensate (see, e.g. \cite{Shly1, Hu}). 
In the temperature range $T_d \gg T \gg  T_{\phi}$, the density fluctuations are suppressed but the phase fluctuations are large, this is 
often refereed to as quasi-condensate (or condensate with fluctuating phase)\cite{Shly1}. In this regime, many experiments \cite{Bess, Bouch} in 1D Bose gases are carried out 
revealing that finite-temperature effects are more pronounced and the properties due to elementary excitations are rather different than those of nearly pure 1D BECs. 

An interesting feature of our formalism is that the noncondensed and the anomalous densities are not independent. 
By deriving an explicit relationship between them, it is possible to eliminate $\tilde{n}$ via \cite{boudj2, boudj5, Cic,Cherny}:
\begin{equation}  \label{Inv}
\tilde{n} = \sqrt{|\tilde{m} |^2+\frac{1}{4}I}-\frac{1}{2}I,
\end{equation}
where $I$ is often known as the Heisenberg invariant \cite{Cic} and represents the variance of the number of noncondensed particles. \\
It is also possible to show that Eq.(\ref {Inv}) can reproduce the mean-field result based on the HFB approximation.
Working in the Bogoliubov quasiparticles space \cite{boudj8} one has 
$\hat a_{\bf k}= u_k \hat b_{\bf k}-v_k \hat b^\dagger_{-\bf k}$, where $\hat b^\dagger_{\bf k}$ and $\hat b_{\bf k}$ are operators of elementary excitations
and $ u_k,v_k$  are the standard Bogoliubov functions. In the quasiparticle vacuum state, $\tilde{n}$ and $\tilde{m}$ may be written as
$\tilde{n}=\sum_k \left[v_k^2+(u_k^2+v_k^2)N_k\right]$ and $\tilde{m}=-\sum_k \left[u_k v_k (2N_k+1)\right]$,
where $N_k=[\exp(\varepsilon_k/T)-1]^{-1}$ are occupation numbers for the excitations.  
Utilizing the orthogonality and symmetry conditions between the functions $u$ and $v$ and using the fact that $2N (x)+1= \coth (x/2)$, we obtain \cite{boudj9}:
\begin{align}\label {heis}
I_k =(2\tilde{n}_k +1)^2-4|\tilde{m}_k |^2 =\text {coth} ^2\left(\frac{\varepsilon_k}{2T}\right).
\end{align}
For an ideal Bose gas where the anomalous density vanishes, $I_k =\text {coth} ^2\left(E_k/2T\right)$ with $E_k$ being the energy of the free particle.
At zero temperature, $I \rightarrow 1$. In the trapped case, the expression of $I$ keeps the same as form Eq.(\ref {heis}) with only replacing the excitation energy
by the energy of the system $\varepsilon_k (x)$ which can be calculated using the semiclassical  approximation.
 
The reduced set of equations are then the coupled equations (\ref{E:td}) with $\tilde{n}$ is replaced by the expression (\ref{Inv}).
This latter clearly shows that $\tilde{m}$ is larger than $\tilde{n}$ at zero temperature, so the neglect of the anomalous density 
is a hazardous approximation.
Importantly, the expression of $\tilde{n}$ not only makes the set (\ref{E:td}) close and enables us to reduce the number of equations making the numerical simulation easier
but also allows us to highlight the role of the anomalous density in the dynamics and the collective modes of the system.

\begin{figure}
  \centering 
  \includegraphics[scale=.55, angle=0]{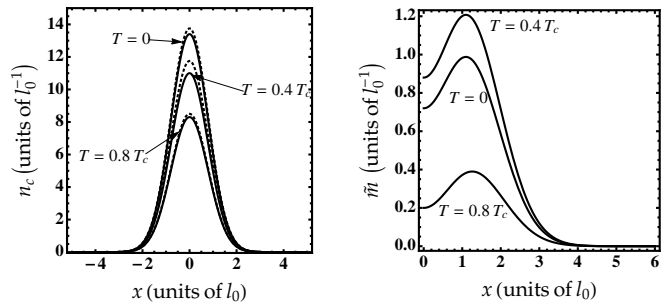}
  \caption{ Density profiles of the condensed (left) and anomalous (right) components 
at different temperatures for $N=25$ and $\gamma_{\text {eff}}=0.01$.
Solid lines: our predictions and dotted lines: HFBP results. Here $T_c^0=T_d/\ln (2N)$ \cite{Kitt} is the
critical temperature for a 1D ideal Bose gas.}
  \label{Prof}
\end{figure}

Inserting Eq.(\ref{Inv}) into set (\ref{E:td}), one obtains 
\begin{subequations}\label{M:td}
\begin{align}
i\hbar \frac{\partial}{\partial t} \Phi &= \left [-\frac{\hbar^2}{2m} \frac{\partial^2}{\partial x^2} +V+g_{ca} n_c  \nonumber  \right. \\ 
&\left. + 2g  \left (\sqrt{|\tilde{m} |^2+\frac{I}{4}}-\frac{I}{2}\right) \right]\Phi, \label{M:td1} \\ 
i\hbar \frac{\partial}{\partial t} \tilde{m}&= 4\left [-\frac{\hbar^2}{2m} \frac{\partial^2}{\partial x^2} +V+2g n_c + G \nonumber \right. \\ 
&\left.  +g_{ac}  \left(\sqrt{|\tilde{m} |^2+\frac{I}{4}}-\frac{I}{2}\right) \right]\tilde{m}, \label{M:td2}
\end{align}
\end{subequations}
where $g_{ac}=2(G+g)$.\\
Remarkably, the coupled TDHFB equations (\ref{M:td}) which describe the dynamics of the condensate and the anomalous density at finite temperature, are 
formally similar to the two coupled time dependent GP equations describing the two components BEC at zero temperature.

To compare our results with the experimental data and previous theoretical treatments, 
it is useful to introduce the effective dimensionless interaction parameter \cite{Choi} instead of $\gamma'$
\begin{equation}  \label{diltparm1}
\gamma_{\text {eff}}= \left[ \frac{2}{ n_{TG} (0) |a_{1D}|} \right]\left(1+\frac {\tilde {m}} {n_c} \right),
\end{equation}
where $n_{TG} =\sqrt{2N m\omega /\hbar }/\pi$ is the analytical TG density in the center of the trap.\\
In what follows, we express lengths, densities and energies in units of the harmonic oscillator length $l_0=\sqrt{\hbar/m \omega}$, $l_0^{-1}$ and $\hbar \omega$, respectively. 

The static equations (\ref{E:td}) or equivalently (\ref{M:td})  and (\ref{diltparm1}) can be solved self-consistently using an appropriate numerical scheme.
In the left panel of Fig.\ref{Prof} we compare our predictions for the condensed density with the HFBP calculation of \cite {Xiao}.
As is clearly seen, there is a definite shift between the two approaches, notably at $T = 0.4 T_c$.
This can be attributed to the presence of the anomalous correlations which reach their maximum at intermediate temperatures 
i.e. $ 0.4 \lesssim T /T_c \lesssim 0.6$ (see the right panel of Fig.\ref{Prof}) \cite {boudj2, Burnet}.
Furthermore, it is quite interesting to observe that the TDHFB predictions become compatible with the HFBP results
as the temperature approaches zero or the transition due to the small contribution of the anomalous density at these points. 
Note that at $T=0$, although the anomalous density is low, it is larger than the noncondensed density \cite {boudj2, Yuk, Griffin, Griffin1, Burnet, boudj3, boudj4};
this is also confirmed by our formula (\ref{Inv}).
Whereas, in the ideal gas limit where $T\geq T_c$, $\tilde{m}$ vanishes  \cite {Griffin,boudj2, boudj3, boudj4} and hence, the TDHFB and the HFBP excellently agree with each other.
We deduce that anomalous fluctuations which arise from the interactions may reduce both the condensed fraction and the critical temperature.
In the TG regime where $\gamma_{\text {eff}} \gg 1$, one can expect that the anomalous density becomes significant.

\section {Collective modes}\label {CM}

The hydrodynamic equations of superfluids have been already successfully employed to predict the collective frequencies of trapped BECs.
In order to study analytically the collective oscillations, 
we write the condensate wavefunction and the anomalous density of the set (\ref {E:td})
in the form\cite{boudj1,boudj5, boudj8}:
\begin{align} \label {eq9}
&\Phi(x,t)=\sqrt{n_c (x,t)} \exp (i \phi(x,t)),  \nonumber \\ 
&\tilde{m} (x,t)=\tilde{m}(x,t) \exp (i \theta(x,t)),
\end{align}
where $\phi$ and $\theta$ are phases of the order parameter and the
anomalous density, respectively. They are real quantities, related to the superfluid and pairing velocities, respectively, as
$v_c=(\hbar/m) \nabla \phi$ and $v_a=(\hbar/m) \nabla \theta$. By substituting expressions
(\ref {eq9}) in Eqs. (\ref {E:td})  and separating real and imaginary parts, one gets the following set of hydrodynamic equations:
\begin{subequations}\label{H:td}
\begin{align}  
&\frac{\partial n_c}{\partial t} +{\bf \nabla}\cdot (n_c v_c)=0, \label{H:td1} \\ 
&\frac{\partial |\tilde{m}|^2} {\partial t} +4{\bf \nabla} \cdot (|\tilde{m}|^2 v_a)=0 \label{H:td2}.
\end{align}
\end{subequations}
Equations (\ref{H:td}) represent equations of continuity expressing the conservation of mass. On the other hand, Euler-like equations read
\begin{subequations}\label{N:td}
\begin{align}  
m\frac{\partial v_c}{\partial t} &=-{\bf \nabla} \bigg [-\frac{\hbar^2}{2m} \frac{\Delta \sqrt{n_c}} {\sqrt{n_c}} +\frac{1}{2} m v_c^2 + V +g_{ca} n_c  \nonumber\\
&+2g\left(\sqrt{|\tilde{m} |^2+\frac{I}{4}}-\frac{I}{2}\right)\bigg], \\ 
m\frac{\partial v_a}{\partial t}& =-4{\bf \nabla} \bigg [-\frac{\hbar^2}{2m} \frac{\Delta \tilde {m}} {\tilde {m}} +\frac{1}{2} m v_a^2 + V +2g n_c +G \nonumber\\
&+g_{ac}\left(\sqrt{|\tilde{m} |^2+\frac{I}{4}}-\frac{I}{2}\right)\bigg] \label{H:td2},
\end{align}
\end{subequations}
where $\Delta \sqrt{n_c}/ \sqrt{n_c}$ and $\Delta\tilde {m}/\tilde {m}$ are quantum pressures associated with condensed and anomalous components, respectively.

\subsection{Analytical results}

For analytical tractability, let us assume that at very low temperature, $I \rightarrow 1$, $\tilde{n} = \sqrt{|\tilde{m} |^2+1/4}-1/2 \simeq \tilde{m}$ 
\cite {boudj2, Yuk, Griffin, Griffin1, Burnet, boudj3, boudj4}.
We then consider small fluctuations of the condensed and anomalous densities $\hat n_c=n_c (x)+\delta \hat n_c$ and $\hat {\tilde{m}}=\tilde{m} (x)+\delta\hat {\tilde{m}}$,
where  $\delta \hat n_c/n_c\ll1$ and $\delta \hat {\tilde{m}}/\tilde{m}\ll1$,  and 
we linearize Eqs.(\ref{H:td}) and (\ref{N:td}) with respect to $\delta \hat n_c$, $\delta\hat {\tilde{m}}$, ${\bf \nabla} \hat \phi$, and ${\bf \nabla} \hat \theta$ 
around the stationary solution. \\
The zero-order terms of these equations give:
\begin{align}  
&\mu_c=-\frac{\hbar^2}{2m} \frac{\Delta \sqrt{n_c}} {\sqrt{n_c}} + V +g_{ca} n_c+2g\tilde {m},  \label{chimB} \\
&\mu_a=4\left[-\frac{\hbar^2}{2m} \frac{\Delta \tilde {m}} {\tilde {m}} + V+g_{ac} \tilde {m}+2g n_c +G\right],  \label{chimm}
\end{align}
where $\mu_a$ is the chemical potential associated with the anomalous density. \\
The first order terms provide equations for the density and phase fluctuations:
\begin{subequations}\label{F:td}
\begin{align}  
 &\hbar \frac{\partial} {\partial t} \frac{\delta \hat n_c} {\sqrt{n_c}} =\left[{\cal L}_c+ g_{ca} n_c+2g\tilde {m} \right] 2\sqrt{n_c} \,\hat \phi,  \label{F:td1} \\ 
&-2\sqrt{n_c}\,\hbar \frac{\partial \hat \phi} {\partial t} =\left[{\cal L}_c +3g_{ca} n_c+2g\tilde {m}\right] \frac{\delta \hat n_c} {\sqrt{n_c}} 
+ 4g\sqrt{n_c} \delta \tilde {m}, \label{F:td2} \\ 
&\frac{\hbar }{4} \frac{\partial} {\partial t} \frac{\delta \tilde {m}} {\tilde {m}}=
\left[{\cal L}_a+g_{ac} \tilde {m}+2g n_c \right] \hat \theta,  \label{F:td3} \\ 
&-\frac{\hbar }{4}\frac{\partial \, \hat \theta}  {\partial t} =
\left[{\cal L}_a+2g_{ac} \tilde {m}+2g n_c\right]  \frac{\delta \tilde {m}} {\tilde {m}} + 2g \delta n_c,  \label{F:td4}  
\end{align}
\end{subequations}
where ${\cal L}_c=-(\hbar^2/2m) \Delta + V -\mu_c$ and ${\cal L}_a=-(\hbar^2/2m) \Delta + V -\mu_a+G$.
Expanding the density and the phase in the basis of the excitations $\varepsilon_j$
\begin{subequations}\label{D:td}
\begin{align}
&\hat \phi(x)=\frac{-i}{2\sqrt{n_c(x)}} \sum_j \left[f_j^{c+}(x) e^{-i\varepsilon_j t/\hbar} \hat {b}_j-h.c.\right], \label{D1:td}\\
&\hat \theta(x)=\frac{-i}{\sqrt{\tilde {m} (x)}} \sum_j \left[f_j^{a+}(x) e^ {-i\varepsilon_j t/\hbar} \hat {b}_j-h.c.\right], \label{D2:td}\\
&\delta \hat n_c (x)=\sqrt{n_c(x)} \sum_j \left[f_j^{c-}(x) e^{-i\varepsilon_j t/\hbar} \hat {b}_j+h.c.\right], \label{D3:td}\\
&\delta \hat {\tilde {m}} (x)=\sqrt{\tilde {m} (x)} \sum_j \left[f_j^{a-}(x)e^{-i\varepsilon_j t/\hbar}  \hat {b}_j+h.c.\right] \label{D4:td},
\end{align}
\end{subequations}
where $f_j^{\pm}=u_j\pm v_j$ satisfy the normalization condition $\int dx [f_j^{+} {f_{j'}^{-}}^*+ f_j^{-} {f_{j'}^{+}}^*]=2\delta_{j,j'}$.
After some algebra, we find the generalized Bogoliubov-de Gennes (BdG) equations:
\begin{subequations}\label{B:td}
\begin{align}  
\varepsilon_j f_j^{c-}(x)&=\left[{\cal L}_c+g_{ca} n_c+2g\tilde {m}\right] f_j^{c+}(x), \label{B1:td} \\
\varepsilon_j f_j^{c+}(x)& =\left[{\cal L}_c+3g_{ca} n_c+2g\tilde {m}\right] f_j^{c-}(x)  \label{B2:td} \\
&+4g\sqrt{\tilde {m} \, n_c}  f_j^{a-}(x), \nonumber  \\  
\varepsilon_j f_j^{a-}(x)&=4\left[{\cal L}_a+g_{ac} \tilde {m}+2g n_c \right] f_j^{a+}(x), \label{B3:td} \\ 
\varepsilon_j f_j^{a+}(x)&= 4\left[{\cal L}_a+2g_{ac} \tilde {m}+2g n_c \right] f_j^{a-}(x) \label{B4:td} \\
&+2g \sqrt{\tilde {m}\,n_c}  f_j^{c-}(x). \nonumber
\end{align}
\end{subequations}
Equations (\ref{B:td}) form a complete set to calculate the ground state and collective modes of  the condensate and the anomalous state at finite temperature.

In a homogeneous case where $V(x)=0$, the spectrum corresponding to the BdG equations (\ref{B:td}) takes the following form
\begin{equation} \label{dis}
\varepsilon_k^{\pm2}= \frac{{\varepsilon_k^c}^2+{\varepsilon_k^a}^2 } {2} \pm \sqrt{\frac{ \left( {\varepsilon_k^c}^2-{\varepsilon_k^a}^2\right)^2 } {4}+ 32 E_k^2 g^2n_c\tilde {m} },
\end{equation} 
where
$\varepsilon_k^c=\sqrt{E_k^2+2g_{ca} n_c E_k }$ and $\varepsilon_k^a=4\sqrt{E_k^2+g_{ac} \tilde{m} E_k }$
denote, respectively the condensate and the anomalous density Bogoliubov dispersions. 
In the long wavelength limit $k \rightarrow 0$,  we have $\varepsilon_k^i= \hbar c_i k$  where $c_i$ ($i=c; a$) is the sound velocity of the condensate and the anomalous component. 
The total dispersion is phonon-like in this limit
\begin{equation} \label{sound}
\varepsilon_k^{\pm}= \hbar c^{\pm} k,
\end{equation} 
with sound velocities $c^{\pm}$  determined by
\begin{equation} \label{sound1}
c^{\pm2} =\frac{1}{2} \left[ {c_c}^2+{c_a}^2 \pm \sqrt{ \left( {c_c}^2-{c_a}^2\right) ^2 + 128 \frac{g^2}{ g_{ca}\, g_{ac} } c_c^2 c_a^2} \right].
\end{equation} 
For $\tilde{m}/ \Phi^2  \in  [-1, (3-\sqrt{17})/4]$, we have $g^2> g_{ca} \,g_{ac}$ and hence, the spectrum (\ref{dis}) becomes unstable. 
In this situation the condensate and the anomalous density spatially separate.

Now we turn to analyzing the behavior of the collective oscillations of a harmonically trapped BEC.

It is obvious that at zero or very low temperatures, the anomalous fraction constitutes the minority component 
and does not affect the condensate which is the majority component.
The system in this case behaves like a highly unbalanced Bose-Bose mixture or a BEC-impurity mixture \cite{boudj5, boudj7}.
In the Thomas-Fermi (TF) approximation, the hydrodynamic equations (\ref{chimB}) and (\ref{chimm}) take the algebraic form
\begin{align}  
& n_c=\frac{\mu_c-m\omega^2 x^2/2}{g_{ca}}, \label{den1} \\ 
&\tilde {m}=\frac{\bar\mu_a- m\bar\omega^2 x^2/2}{g_{ac}} \label{den2},
\end{align}
where $\bar\mu_a=\mu_a(1/4- 2\alpha \mu_c /\mu_a )$, $\bar\omega^2=\omega^2(1-2\alpha)$, and $\alpha=g / g_{ca}$.
In Eq.(\ref{den2}), we have omitted the term $G/g_{ac}$ since its contribution is negligible at temperatures tending to zero. \\
The condensate evidently behaves as being unique and its dispersion relation in the quasi-1D case  can be obtained by
considering low-energy excitations $(\varepsilon_j\ll /\mu_c)$. 
Using  the TF equation (\ref{den1}) and substituting $f_j^{c-}$ from Eq.(\ref{B1:td}) into Eq.(\ref{B2:td}),
we then obtain an equation for $f_j^{c+}$. 
The solution of such an equation gives the frequencies $\varepsilon_j^c =\hbar \omega \sqrt{j (j+1)/2}$ \cite{Menot, Shly1}, 
where the quantum number $j$ is an integer and depends on the mode of interest. 
For the center-of-mass (dipole) mode $(j = 1)$, the condensate has simply $\varepsilon^c =\hbar\omega $, while
for the breathing mode $(j = 2)$, $\varepsilon^c =\sqrt{3}\hbar \omega $ which well coincides with collective modes of a purely BEC at zero temperature.

To calculate the collective modes of the anomalous fraction which is the minority component, we suppose that for the low frequency mode the spatial
variation of the condensate component is just a higher order correction, i.e. $ {\bf \nabla} \delta \tilde{m} \gg {\bf \nabla} \delta {n_c} $. 
In this case the dispersion relation of $\tilde{m}$ takes the form:
\begin{equation}  \label{disa}
\varepsilon_j^a =4\hbar \bar\omega \sqrt{j (j+1)/2}.
\end{equation}
The relation (\ref{disa}) shows that for all the repulsive coupling constants the dynamics of the anomalous fraction is faster in the presence of the condensate.
Indeed, this is natural since the anomalous density itself is related to the thermal cloud which acquires higher energy.
For the dipole and the breathing modes, $\varepsilon_j^a$ has, respectively $\varepsilon^a =4\hbar\omega \sqrt{1/2-\alpha}$ and 
$\varepsilon^a =4\sqrt{3}\hbar\omega \sqrt{1/2-\alpha}$.
An unexpected result is that the dispersion of the anomalous fraction depends explicitly on the interaction strength
which is in disagreement with the above spectrum of the condensate \cite{Menot, Shly1}.\\
The wavefunctions associated with the anomalous component can be given as
\begin{equation}  \label{WFA}
f_j^{a \pm}(y_a) = \left(\frac{j+1/2}{L_{TF}^{a}} \right)^{ 1/2} \left[\frac{\bar\mu_a}{\varepsilon_j^a} \left(1-y_a^2\right)\right]^{\pm 1/2} P_j(y_a),
\end{equation}
where $P_j$ are Legendre polynomials, $y_a=x/L_{TF}^{a}$, and $L_{TF}^{a}=\sqrt{2\bar\mu_a/m\bar\omega^2}$ is the TF size of the anomalous component.

\begin{figure}
  \centering 
  \includegraphics[scale=.8, angle=0]{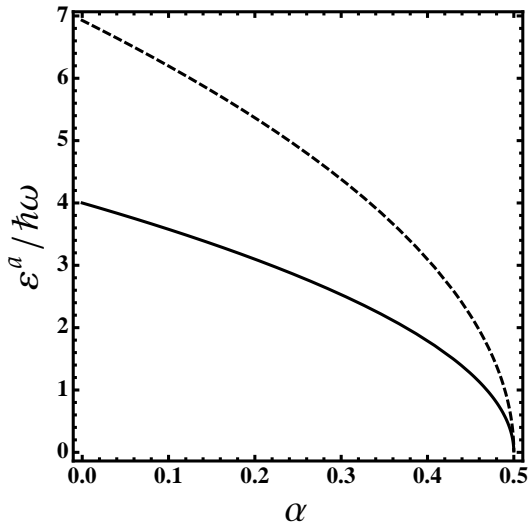}
  \caption{ Center-of-mass (solid lines) and breathing (dashed lines) modes of the anomalous component as a function of $\alpha$.}
  \label{Mod}
\end{figure}

Figure \ref{Mod} depicts that the center-of-mass and the breathing modes of the anomalous component decrease with increasing $\alpha$. 
From $\alpha > 0.5$,  $\varepsilon_j^a$ vanishes since the anomalous density disappears and thus, the condensate survives alone as we have foreseen above.

The mean-square fluctuations of the anomalous density is defined at equal times as 
$\langle (\Delta\hat {\tilde {m}} (y) )^2\rangle=\langle [\delta\hat {\tilde {m}} (y)-\delta\hat {\tilde {m}} (0)]^2\rangle$. 
A straightforward calculation using  Eqs.(\ref{D4:td}) and (\ref{WFA}) yields
\begin{align}   \label{DF}
\frac {\langle (\Delta\hat {\tilde {m}} (y_a) )^2\rangle}{\tilde{m}^2 (0)} &= \sum_j \frac{\varepsilon_j^a (j+1/2)}{\bar\mu_a L^a_{TF} \tilde{m} (0)}  \left[P_j(y_a) -P_j(0)\right]^2 
\nonumber \\
&\times \coth \left(\varepsilon_j^a/2T \right), 
\end{align}  
where $\tilde{m} (0)$ is the central density which can be computed via Eq.(\ref{den2}).
At low temperature, Eq. (\ref{DF}) becomes
\begin{equation}   \label{DF1}
\frac {\langle (\Delta\hat {\tilde {m}} (y_a) )^2\rangle}{\tilde{m}^2 (0)} = \frac{T}{\bar\mu_a} \sum_j \frac{ 2j+1}{L^a_{TF} \tilde{m} (0)}  \left[P_j(y_a) -P_j(0)\right]^2. 
\end{equation}
Equation (\ref{DF1}) shows that the anomalous density fluctuations are strongly suppressed at temperatures $T\ll \bar\mu_a$.
This confirms the existence of the anomalous density in a trapped 1D system as we have anticipated in the introduction.
At higher temperatures, the main contribution to the density fluctuations (\ref{DF}) comes from quasiclassical
excitations $(j \gg 1)$. So, the fluctuations are small: $\langle (\Delta\hat {\tilde {m}} (y) )^2\rangle/\tilde{m}^2 (0) \sim 1/(\tilde {m} (0) \xi_a)$ 
with $\xi_a=\hbar/\sqrt{m \tilde {m} (0) g_{ac}}$ being the correlation length of the anomalous component.

\subsection{Numerical simulations}

In this subsection, we investigate more precisely the effects of the anomalous density on the collective mode frequencies of the condensate at finite temperature.
To this end, we solve numerically the full BdG equations which can be obtained easily from Eqs.(\ref{E:td}) and (\ref{Inv})
and compare our findings with existing previous analytical, numerical and experimental predictions.

\begin{figure}
  \centering {
  \includegraphics[scale=.8, angle=0]{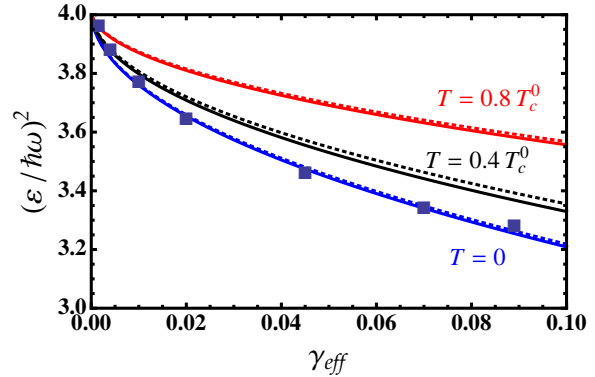}}
  \caption{ (Color online)  Breathing mode frequency as a function of $\gamma_{\text {eff}} $ for $N=25$ in the weakly-interacting regime.
Solid line : TDHFB predictions. Dotted line : HFBP results. Squares: Generalized GP equation at $T=0$ of Refs \cite{Choi, Xiao}.}
  \label{BMT}
\end{figure}

Figure \ref{BMT} depicts that the TDHFB excitation frequencies slightly diverge from the generalized GP and the HFBP
predictions at $T=0$. 
At intermediate temperatures, $T=0.4T_c$, the discrepancy between the TDHFB and the HFBP approaches becomes perceptible $\sim 4\%$ at  $\gamma_{\text {eff}}=0.1$
owing to the strong contribution of the anomalous fluctuations.
We can see also that the shift rises with rising $\gamma_{\text {eff}}$ and is downward in frequency.
Note that this behavior holds also in a trapped 3D Bose gas \cite{Burnet}.
In the same figure, we observe that as the critical temperature is approached ($T \simeq T_c$), 
the frequency of the breathing mode increases until it reaches its ideal gas value $\varepsilon/\hbar \omega =2$, 
in good agreement with the recent theoretical results of Ref.\cite{Xiao}.

\begin{figure}
  \centering {
  \includegraphics[scale=.8, angle=0]{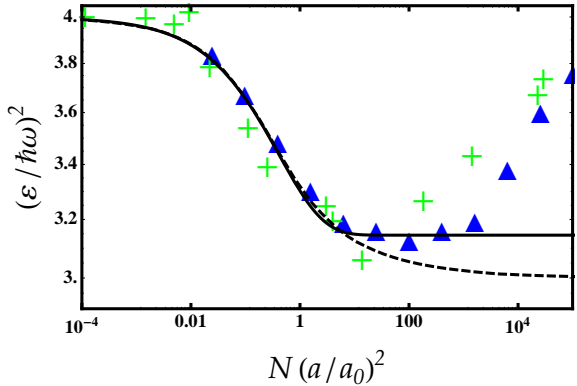}}
  \caption{ (Color online)  Breathing mode frequency as a function of interaction strength $N (a/a_0)^2$ in logarithmic scale.
Solid line : TDHFB predictions. Dashed line : LDA results with $N=25$ \cite{Menot}. Filled triangles : Monte carlo simulation with $N$=25 \cite {Zvon}. 
Plus: Innsbruck experiment \cite{Haller}. }
  \label{BM}
\end{figure}

In Fig.\ref{BM}  we plot our predictions  for the breathing mode frequency as a function of $N (a/a_0)^2$ and compare the results with 
the experimentally measured frequencies \cite{Haller}, Monte carlo simulation points \cite{Zvon} and the local density approximation (LDA) \cite{Menot}.
For $N (a/a_0)^2 \lesssim 100$, our predictions are in qualititaive agreement with Innsbruck expreiment \cite{Haller} and Monte carlo simulation \cite{Zvon}.
Whereas for large $N (a/a_0)^2$, frequencies of both experimental and numerical data are higher than our theory predicts.
The TDHFB and  the LDA curves \cite{Menot} agree for  $N (a/a_0)^2 < 1$.
This match proves that the anomalous correlations have a minor role in the limit of very weak interactions.
Note that a recent finite temperature investigation within a variational approach has shown the appearance of a broad minimum in the mode frequency,
with its position shifting to the strong regime when temperature increases \cite {Hu}.
The present TDHFB theory is not applicable beyond $N (a/a_0)^2 >100$, and therefore the regime where the broad temperature-dependent minimum 
in the mode frequency occurs can not be reproduced here.

\begin{figure}
  \centering {
  \includegraphics[scale=.8, angle=0]{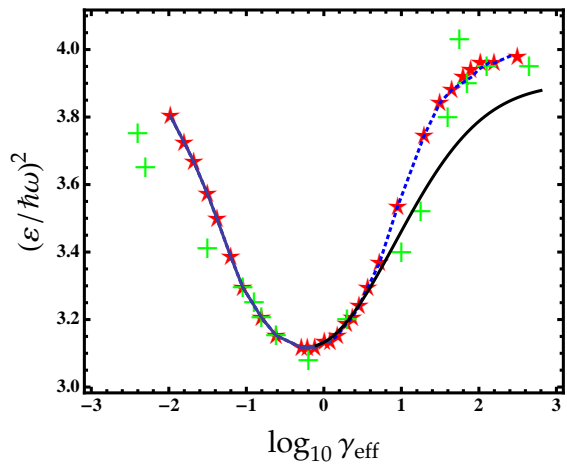}}
  \caption{ (Color online)  Breathing mode frequency as a function of $\text {log}_{10} \gamma_{\text {eff}} $.
Solid black line : TDHFB predictions.  Blue  dotted line: Generalized GP equation \cite{Xiao}. Stars: MNLSE results \cite{Choi}.
 Plus: Innsbruck experiment \cite{Haller}.}
  \label{BMG}
\end{figure}

Figure \ref{BMG} displays the behavior of the breathing modes in terms of the gas paprameter $\gamma_{\text {eff}} $.
It is clearly visible that the curves of the TDHFB approach, MNLSE \cite{Choi} and generalized GP equation \cite{Xiao} agree with each other,  
while all of these theories cannot reproduce the experimentally measured mode frequency \cite{Haller} in the spectrum of the non-interacting regime.
A careful analysis of the same figure shows that our TDHFB theory improves the previous theoretical treatments in the region
$ 0<\text {log}_{10} \gamma_{\text {eff}} \leq 1$.  This correction brings our theory in qualitative concordance with Innsbruck experiment data \cite{Haller} in the weakly interacting regime.
For $\gamma_{\text {eff}} \gg1$ where the system reaches the TG limit,  the TDHFB theory fails to capture it.
A source of this discrepancy  can be,  as we have stated above, the considerable effects of the anomalous density which causes a huge loss
of atoms during the oscillation even at zero temperature.

\section{Conclusion} \label{concl}

We have studied the finite-temperature collective modes of a trapped atomic Bose gas in a quasi-1D geometry using the TDHFB theory.
This allows us to analyze the coupled dynamics of the condensed and the anomalous components which offers straightforward implications for interpretation of recent experiments. 
We have presented detailed results for the condensed and anomalous densities in terms of temperature. 

Within the realm of the hydrodynamic approach, we have derived analytical expressions for the spectrum of the condensate and the anomalous components.  
Surprisingly, we found that the dipole and the breathing modes of the anomalous fraction are enhanced by increasing the inter-component interaction.
The density fluctuations of the anomalous component corresponding to these modes has been found to be small at temperatures $T\ll \bar\mu_a$.

By fully self-consistent numerical calculation of the TDHFB-BdG equations, we have determined the temperature dependence of the excitations of the system.
We have shown that the presence of the anomalous correlations leads to shift the calculated breathing mode frequency from those obtained by the HFBP 
approximation, in particular at intermediate temperatures. The height of the shift increases as the effective interaction parameter increases.
At zero temperature, we have compared our findings for the breathing mode frequency with recent experimental data, Monte carlo simulations and previous theoretical predictions.
Results show that although TDHFB predictions, Monte carlo calculations, and previous theoretical theories are in very good agreement with each other,
they conflict with measured values in the noninteracting regime. 
In contrast, the TDHFB results slightly correct the existing previous theoretical works in the mean-field regime, making  
the theory in qualitative agreement with the recent experiment \cite{Haller}, while it fails
in the strongly interacting limit due to the indispensable role of the anomalous correlations which render the system highly correlated.

Clearly, further works, both experimentally and theoretically, will be required to solve this issue.
One way of grafting the desired corrections is to fully include the dynamics of higher order quantum fluctuations.

%On the other hand, we have explored the behavior of solitons in the BEC-Pairing  mixture. Our results show the spontaneous generation of bright/dark solitons
%in the pairing component without any squeezing of the trapping potential. 
%The size of these solitons may be adjusted by varying the inter-component interaction strength $\alpha$.
%Furthermore, for sufficiently high pairing instability, the soliton-pair grows and form a soliton-molecule with the condensed soliton at a certain separation distance.
%The pairing instability may lead also to split the soliton in the condensed component into several peaks forming a soliton-train.
%For very strong anomalous fluctuations, the system becomes unstable and hence, our mean field TDHFB is no longer valid.

\section{Acknowledgements}
We are indebted to Mikhail Zvonarev, Stephen Choi, Maxim Olshanii, Isabelle Bouchoule, Hanns-Christoph Naegerl
and Xiao-Long Chen for fruitful discussion and for giving us the numerical and the experimental data.
%This work was supported by the Algerian government under Research Grant No. CNEPRU-B00L02UN020120130004. 

\end{document}